%% file: main.tex
\documentclass[a4paper, amsfonts, superscriptaddress, amssymb, amsmath, reprint, showkeys, nofootinbib, twoside]{revtex4-1}
\usepackage[english]{babel}
\usepackage[utf8]{inputenc}
\usepackage[colorinlistoftodos, color=green!40, prependcaption]{todonotes}
\input{preamble}
\usepackage[pdftex, pdftitle={Article}, pdfauthor={Author}]{hyperref} 
\begin{document}

\title{Molecular electric dipole moments: from light to heavy molecules using a relativistic VQE algorithm}

\author{K. R. Swain}
    \email{swain1704@gmail.com}
    \affiliation{Centre for Quantum Engineering, Research and Education, TCG Crest, Kolkata 700091, India}
\author{V. S. Prasannaa}
    \email{srinivasaprasannaa@gmail.com}
    \affiliation{Centre for Quantum Engineering, Research and Education, TCG Crest, Kolkata 700091, India}
\author{Kenji Sugisaki}
    \email{ksugisaki@keio.jp}
    \affiliation{Centre for Quantum Engineering, Research and Education, TCG Crest, Kolkata 700091, India}
    \affiliation{Graduate School of Science and Technology, Keio University, 7-7 Shinkawasaki, Saiwai-ku, Kawasaki, Kanagawa 212-0032, Japan.}
    \affiliation{Keio University Quantum Computing Center, 3-14-1 Hiyoshi, Kohoku-ku, Yokohama, Kanagawa 223-8522, Japan.}
\author{B. P. Das}
    \email{bpdas.iia@gmail.com}
    \affiliation{Centre for Quantum Engineering, Research and Education, TCG Crest, Kolkata 700091, India}
    \affiliation{Department of Physics, Tokyo Institute of Technology, 2-12-1 Ookayama, Meguro-ku, Japan}

\date{\today} 

\begin{abstract}
The quantum-classical hybrid Variational Quantum Eigensolver (VQE) algorithm is recognized to be the most suitable approach to obtain ground state energies of quantum many-body systems in the noisy intermediate scale quantum era. In this work, we extend the VQE algorithm to the relativistic regime and carry out quantum simulations to obtain ground state energies as well as molecular permanent electric dipole moments of single-valence diatomic molecules, beginning with the light BeH molecule and all the way to the heavy radioactive RaH molecule. We study the correlation trends in these systems as well as assess the precision in our results within our active space of 12 qubits.

\end{abstract}


\maketitle

\input{sections/section01.tex}  
\input{sections/section02.tex}
\input{sections/section03.tex}
\input{sections/section04.tex}
\input{sections/acknowledgements.tex}

\bibliography{ref1}

\clearpage
\appendix
\input{sections/appendix1.tex}

\end{document}

%% file: preamble.tex
\usepackage{amsthm}
\usepackage{mathtools}
\usepackage{physics}
\usepackage{xcolor}
\usepackage{graphicx}
\usepackage[left=23mm,right=13mm,top=35mm,columnsep=15pt]{geometry} 
\usepackage{adjustbox}
\usepackage{placeins}
\usepackage[T1]{fontenc}
\usepackage{lipsum}
\usepackage{csquotes}
\usepackage{float}

\usepackage{physics}
\usepackage{braket}
\usepackage{esvect}
\usepackage{amsmath}

%% file: sections/section01.tex
\section{Introduction} \label{sec:outline}

Since Feynman’s proposal in 1982 \cite{c5}, there have been many advances in the field of quantum computing. A significant amount of effort has gone into developing quantum algorithms for the simulation of quantum many-body systems \cite{c21, c22, c23,I3}. These algorithms have the potential for achieving quantum advantage in atomic and molecular calculations on quantum computers, after resource optimization and quantum error correction/mitigation. Even with the state-of-the-art traditional supercomputers, this problem is very challenging, since the computational cost needed to describe atomic and molecular states grows exponentially with the system size. The wavefunction of an atom or molecule is entangled, and this property can be exploited by quantum devices to achieve an exponential speedup over conventional computers \cite{HUANG20051, kais2, Rongxin}.
 
The quantum phase estimation algorithm \cite{I1,I2} was proposed to efficiently solve atomic and molecular problems on a quantum computer. However, since even for medium-sized systems, this algorithm requires many gates (and hence leading to deep quantum circuits), the approach is strongly limited by noise and decoherence effects that are prevalent in the current noisy intermediate scale quantum (NISQ) era quantum devices. However, the classical-quantum hybrid variational quantum eigensolver (VQE) algorithm addresses this issue with relatively shallow circuits, thus allowing it to be run on the available NISQ \cite{c6} devices. 

Thus far, the VQE algorithm has predominantly been employed to calculate ground state energies and first ionization energies of atomic, molecular and ionic systems \cite{m5, m6, m30, m31, N1, N2, N3}. It has also been experimentally realized on different quantum hardware platforms \cite{I8, I6, I7, I5, N10, N11, N12, m7, m8, m9, m10, m11}. However, there are not many works that calculate atomic and molecular properties besides energies \cite{PhysRevA.105.012425, PhysRevLett.122.230401, m30}. Further, to the best of our knowledge, only one work treats relativistic effects in such systems \cite{N5} within the framework of the VQE algorithm. There are a few studies that incorporate relativistic effects in the context of other quantum computing algorithms \cite{rq, rs}.

In this work, we calculate the permanent electric dipole moments (PDMs) of single valence molecular systems along with their ground state energies, by implementing the VQE algorithm in a relativistic framework. Our choice spans the single-valence alkaline earth metal monohydrides, including the laser cooled BaH \cite{McNally_2020} and the radioactive and heavy RaH molecule, where relativistic effects become progressively important.

We discuss the theory and the adopted methodology in our work, in Section II. This is followed by Section III, where we show the results obtained for the ground state energies and PDMs of the chosen molecules using the relativistic VQE algorithm. We finally conclude in Section IV. 

We expect that our work would be a first step in determining the PDMs of more complex molecules, such as SrF and BaF \cite{I9}, which are laser coolable and can be trapped in optical lattices \cite{I10}. The PDM itself is a useful property and helps one to determine, for example, the dipole-dipole interaction in the low-temperature regime \cite{Trefzger_2011}, which can give rise to novel quantum states of matter like the supersolid state \cite{PhysRevA.80.043614}, thereby enabling one to understand some fundamental aspects of condensed matter physics. However, our work also opens new avenues in exploring several other atomic and molecular properties of interest to various applications such as new physics beyond standard model of elementary particles, where the effects of relativity become prominent \cite{BSM1, BSM2}. 

%% file: sections/section02.tex
\section{THEORY AND METHODOLOGY} \label{sec:develop}

For a given many-body Hamiltonian $\hat{H}$, and a trial wave function $\ket{\Psi(\theta)}$ with variational parameters $\{\theta\}$ $\in$ $\{\theta_1,\theta_2, \ldots\}$, minimizing the functional, $E(\theta)$, with respect to $\{\theta\}$ yields an energy that is an upper bound to the true ground state value, $E_0$. Mathematically, we can express it as
\begin{equation}
          E(\theta) = \frac{\expval{\hat{H}}{\Psi(\theta)}}{\braket{\Psi(\theta)|\Psi(\theta)}} \geq  E_0. 
\end{equation}

The above statement conveys the essence of the Rayleigh-Ritz variational principle \cite{T1}. The trial wavefunction, $\ket{\Psi(\theta)}$, also called the ansatz, should be realizable as a quantum circuit so that it can be executed on a quantum device. In our work, we choose our ansatz to be the unitary coupled cluster (UCC) ansatz \cite{T4, T5}, which is given as
\begin{equation}
\begin{split}
        \ket{\Psi(\theta)} & = \hat{U}(\theta) \ket{\Phi_0}\\
        & = e^{\hat{T}-\hat{T}^\dagger} \ket{\Phi_0}\\
        & = e^{\hat{\tau}} \ket{\Phi_0},
\end{split}
\end{equation}
where $\ket{\Phi_0}$ is the initial/reference Hartree-Fock (HF) state, and $\hat{U}(\theta)$ is the UCC operator. $\hat{T}$ and $\hat{T}^\dagger$ are the fermionic excitation and de-excitation operators respectively. In our work, we have only considered single (S) and double (D) excitations, and hence in our UCC operator, $e^{\hat{\tau}}=e^{\hat{\tau_1}+{\hat{\tau_2}}}$  (called as UCCSD ansatz). Further, $\hat{T_1}=\sum_{i,a} \theta^a_i \hat{a}_a^\dagger \hat{a}_i$ and $\hat{T_2} = \sum_{i,j,a,b} \theta^{ab}_{ij} \hat{a}_a^\dagger \hat{a}_b^\dagger \hat{a}_j \hat{a}_i$, where $\hat{a}_a^\dagger$ and $\hat{a}_i$ are creation and annihilation operators, which create and annihilate an electron in spin orbitals $a$ and $i$ respectively. We have used the subscripts i and j for occupied, a and b for unoccupied, and p, q, r, and s for general spin orbitals throughout the manuscript. The energy functional in eqn (1) now becomes
\begin{equation}
        E(\theta) = \expval{e^{\hat{\tau}^\dagger(\theta)}\hat{H} e^{\hat{\tau}(\theta)}}{{\Phi_0}}.
\end{equation}

It is worth mentioning at this point that approaches based on the coupled cluster method have been found to be most suitable for calculations of atomic and molecular properties \cite{T7, T8}. In the UCC approach, the usual CC method is modified to be both variational and unitary, and therefore it is amenable for implementation on a quantum computer while retaining the predictive ability of the traditional CC method.

    
The molecular many-body Hamiltonian, $\hat{H}$, can be written in the second quantized notation as 
\begin{equation}
                    \hat{H} = \sum_{p,q} h_{pq} \hat{a}_p^\dagger \hat{a}_q + \frac{1}{2} \sum_{p,q,r,s} v_{pqrs} \hat{a}_p^\dagger\hat{a}_q^\dagger \hat{a_s}\hat{a_r}, 
\end{equation}
The summation is over the number of spin orbitals, which depends on the choice of the employed single particle basis set. For the non-relativistic calculations, the one- and two-electron integrals $h_{pq}$ and $v_{pqrs}$, are evaluated on a traditional computer using a  suitable program, for example, the Pyscf \cite{T2} package.
    
We use the Jordan–Wigner (JW) mapping \cite{Jordan1928, Tranter2018} to transform the second quantized operators in eqn (4) to tensor products of Pauli operators. The Hamiltonian in eqn (4) now takes the form
\begin{equation}
        \hat{H}'=\sum_{j}^{M} \alpha_j \hat{P}_j,
\end{equation}
where $\alpha_j$ are real scalar coefficients which depend on $h_{pq}$ and $v_{pqrs}$. $P_j$ are Pauli strings represented by tensor products of Pauli operators $\{I,X,Y,Z\}$ and M denotes the number of Pauli strings in the Hamiltonian. Now, eqn (3) can be written as
\begin{equation}
        E(\theta) = \sum_{j}^{M} \alpha_j \expval{U'^{\dagger}(\theta) \hat{P}_j U'(\theta) }{\Phi'_0}. 
\end{equation}
where $U'$ and $\ket{{\Phi'_0}}$ refer to the JW-transformed U and $\ket{{\Phi_0}}$ respectively. Each term in eqn (6) corresponds to the expectation value of a Pauli string $\hat{P}_j$, which will be evaluated on a quantum computer, while the summation and energy minimization will be done on a classical computer. The energy is minimized by updating the parameters in an iterative process using an appropriate optimization algorithm. In this work, we begin with zero initial parameters and employ the SLSQP classical optimizer \cite{T6} to obtain the optimal VQE parameters. 
    
For the relativistic calculations, we begin with the Dirac-Coulomb Hamiltonian (DCH), which is given by 
\begin{equation}
            \hat{H}_{DC}= \sum_i (c \vv{\alpha} \cdot \vv{p_i}+ \beta m_i c^2) +\sum_i V_{nuc}(r_i) + \frac{1}{2} \sum_{i \neq j} \frac{e^2}{r_{ij}}, 
\end{equation}
where $\vv{\alpha} =\begin{pmatrix}
                        0 & \vv{\sigma} \\
                        \vv{\sigma} & 0
                     \end{pmatrix}$, 
    $\beta =\begin{pmatrix}
                        \mathbb{I} & 0 \\
                        0 & -\mathbb{I}
                             \end{pmatrix}$, 
$\vv{\sigma}$ are the Pauli matrices and $\mathbb{I}$ is the $(2\times2)$ identity matrix. When the DCH is converted to its second quantized form, $h_{pq}$ and $v_{pqrs}$ in eqn (4) are obtained using a suitable relativistic code. $\ket{\Phi_0}$ from eqn (2) is the Dirac-Hartree-Fock (DF) state. It is important to stress that since the wavefunction is a four-component spinor, there are more integrals to evaluate in order to arrive at the value of a given $h_{pq}$, for example. This is because a given $h_{pq}$ is a sum of integrals of types LL,LS, SL, and SS, where LL means that the bra and the ket have the large components of atomic orbitals, LS means the large component in the bra and small component in the ket, etc. However, although more number of integrals are calculated in this way, we only feed into VQE a single number for a given $h_{pq}$ or for that matter, a given $v_{pqrs}$. We add that the small components are generated from the large components via restricted kinetic balance condition \cite{Cramer2008}.

We use the relativistic integrals generated by DIRAC22 \cite{T3}. We note that throughout this work, even for non-relativistic (NR) calculations, we employ the DIRAC22 code for generating the NR Hamiltonian integrals. We then carry out a NR/relativistic UCCSD-VQE computation to obtain the ground state energy by employing the statevector backend. The PDM of a diatomic molecule can be obtained by $ \braket{\hat{D}} = \expval{\hat{D}}{\Psi(\theta_c)}$; where $\ket{\Psi(\theta_c)}$ is the ground state wavefunction which is constructed from the converged amplitudes (denoted by the subscript, `$c$') of the VQE algorithm. The dipole moment operator, $\hat{D}$, contains two terms, nuclear and electronic. Here, the nuclear part contains only one term, since we have chosen one of the two atoms in the diatomic molecule to be positioned at the origin. Therefore, within the Born-Oppenheimer approximation, $\hat{D}$ is  expressed as
\begin{equation}
        \hat{D} = e \left(Z_N R_N -\sum_j r_j \right), 
\end{equation}
    
\noindent where $e$ is the electronic charge, $r_j$ is the position coordinate of the $j^{th}$ electron from the origin, $R_N$ is the bond length of the molecule, and $Z_N$ is the nuclear charge of the second atom which is not at the origin. We obtained both the non-relativistic and relativistic dipole integrals from the DIRAC22 code and calculated the dipole moment of the diatomic molecules by taking the expectation value of the dipole operator with respect to the ground state wavefunction.

%% file: sections/section03.tex
\section{Results and discussion} \label{sec:results}
\subsection{Details of computation}
This section presents and discusses our results for the PDMs and ground state energies of the LiH, BeH, MgH, CaH, SrH, BaH, and RaH molecules in the contracted dyall.v2z basis sets \cite{dyall}. Except LiH, all the other considered molecules contain one unpaired electron (single valence molecules). To that end, we have made appropriate non-trivial changes to Qiskit 0.26 \cite{m30} to adapt the code to handle such systems. Further details can be found in reference \cite{R4}. We choose the following values for the equilibrium bond lengths (in \AA): LiH : 1.595  \cite{R1}, BeH : 1.342 \cite{R3}, MgH : 1.7297 \cite{R3}, CaH : 2.0025 \cite{R3}, SrH : 2.1461 \cite{R3}, BaH : 2.2319 \cite{R3}, RaH : 2.43 \cite{RaH}. Further, our active space for LiH is built out of four occupied spin orbitals (s.o.) and eight unoccupied ones, while for the rest of the systems (which are single valence), it comprises of five occupied s.o. and seven unoccupied ones. 

    \begin{figure}[!t]
        \centering
        \includegraphics[width=0.47\textwidth]{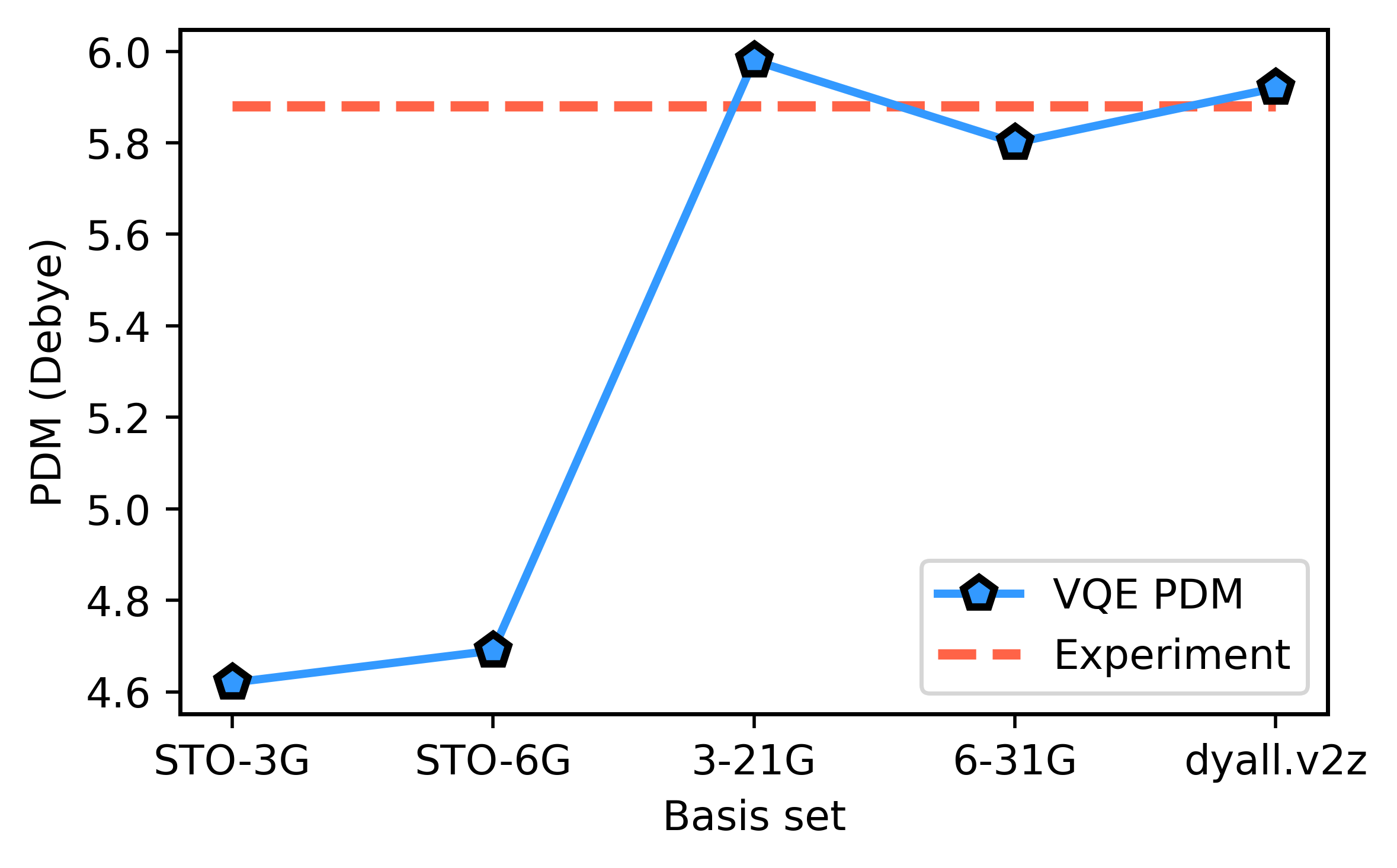}
        \caption{Figure illustrating the behaviour of the PDM, of the LiH molecule obtained using VQE algorithm, with different basis sets.}
        \label{fig:1}
    \end{figure}

\subsection{Choice of basis : analysis}
We now comment on our choice of basis sets. A molecular orbital is expanded as a linear combination of atomic orbitals (basis functions). In the minimal STO-3G/STO-6G basis, each basis function is built out of a linear combination of three/six Gaussians in order to mimic a Staler-type function. One may expect that the quality of results will improve, when more than one basis function is used for each valence atomic orbital. In the 3-21G/6-31G basis set, the core orbitals are each built out of one basis function (a linear combination of three/six Gaussians), whereas in the two valence orbitals, the first is composed of two/three basis functions and the second is composed of one. 3-21G is also called a valence double-zeta (dz) basis set because it uses two basis functions to describe one valence atomic orbital. One must be very careful when selecting the basis sets for heavier elements in relativistic calculations. For the Dirac-Coulomb calculations, which we perform in our work, Dyall and a number of other groups have prepared high quality correlation consistent dz basis, which include correlating functions for the valence and outer core orbitals. In dyall.v2z basis, ‘2z’ denotes double zeta, whereas ‘v’ stands for valence basis sets.
    
Figure \ref{fig:1} shows the values of the PDM of LiH obtained using the VQE algorithm in five different basis sets. Each data point is the result of a twelve qubit calculation. It can be seen from the figure that Dyall's v2z basis works best among the other basis sets as the PDM obtained using it is closer to the experimental value. Hence, we perform all the calculations hereafter in Dyall's v2z basis \cite{Dyall2016}.

We now proceed to investigate the interplay of relativistic and correlation effects through our theoretical studies of the ground state energies and PDMs of the considered molecules. In order to do so, we perform six calculations per molecule: mean-field (HF and DF), quantum-classical hybrid VQE algorithm (our NR and relativistic (rel) codes), and complete active space configuration interaction (CASCI; best possible result within a single particle basis) in the NR and rel regimes. We first focus on the contributions from the relativistic effects, followed by those from correlation effects. We can determine the relativistic effects of the desired molecule by comparing the HF and DF values, or VQE NR and rel results from our codes, or CASCI NR and rel results. We also give percentage fraction differences between the HF and CASCI, as well as the HF and VQE methods, to get an estimate of the correlation effects that have been captured by the VQE method. After determining the relativistic and correlation contributions, we examine the ‘net’ effect due to their interplay, resulting from cancellations. To test the proper implementation of our code, we compared the VQE results between the Qiskit NR and our NR methods for smaller basis sets. As the results match exactly up to six decimal places in energy and second decimal place in the PDM, we have not included the results explicitly. All our calculated results are presented in Table \ref{tableS1} of the supplementary material. 
    \begin{figure}
        \centering
        \includegraphics[width=0.48\textwidth]{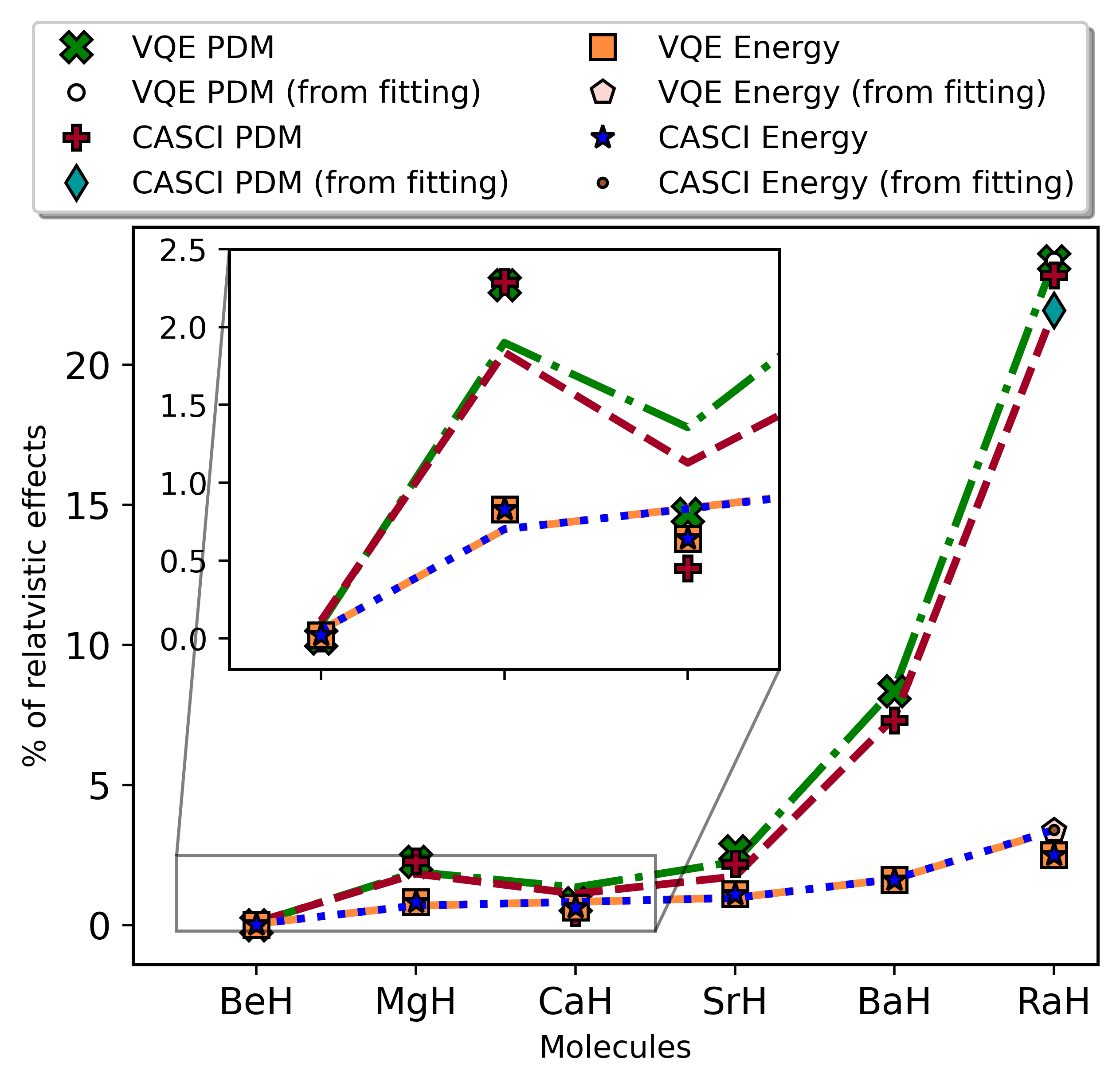}
        \caption{Figure illustrating the \% rel. effects = $\frac{rel-NR}{NR} \times 100 $ across the alkaline earth metal monohydride systems. It also shows the agreement between the actual calculated VQE values of energy and PDM with a third order polynomial fit, for the heavy radioactive RaH molecule.}
        \label{fig:2}
    \end{figure}

    \subsection{Relativistic effects}
We begin by examining our results for the size of relativistic effects in the ground state energies and PDMs of the considered systems. For the case of LiH, as expected, relativistic effects are not noticeable. They contribute 0.81 mHa to the ground state energy at the mean field level, while they account for 0.80 mHa to the energy for both VQE and CASCI. This translates to about 0.01 $\%$  in all three cases. For the alkaline earth metal monohydrides, we observe that the trends in energies follow the expected pattern of relativistic effects increasing in importance as we progress from the lighter to the heavier systems. The lowest contribution from relativity at the mean field level is found to be 2.8 mHa for BeH, whereas in contrast, we find the effect to be 121.71 mHa for RaH. After the inclusion of correlation effects, we find that both for VQE and CASCI, the importance of relativity ranges from 2.7 mHa for BeH to 121.66 mHa (for VQE) and 121.09 mHa (for CASCI) for RaH. These numbers correspond to 0.02 $\%$  for BeH to about 2.50 $\%$ for RaH relative to the NR values. We now focus on the analysis of the precision of our results. The percentage fraction difference (PFD) results for VQE deviate from those of CASCI by 0.91 $\%$ (for the SrH molecule), while it is perfect agreement for all other molecules. 

We now report the observed trends in the PDM. We find that for LiH, there is practically no difference between the NR and the relativistic values of the PDM. For the alkaline earth metal monohydride molecules, relativistic effects are noticeable clearly from the computations of the PDM as we go towards heavier molecules. We see immediately from Figure \ref{fig:2} that the PDM is substantially more sensitive to relativistic effects than energy. For example, the percentage fraction difference of relativistic to non-relativistic ground state energy for RaH is 2.50, while it is 9.56 times that value for the PDM. However, as we approach towards heavier molecules, there is a certain degree of disagreement in the PDM between VQE and CASCI methods. The missing excitations in the former may be the cause of the discrepancy. This could be attributed to the dominance of different correlation effects in the two different properties that we have considered \cite{R3}. 

Figure \ref{fig:2} also shows the global trends that we observe in more detail. A testament to the performance of the relativistic VQE algorithm would be to check if an extrapolation to RaH, using the data up to BaH, would be sufficiently close to the actual VQE computation. We carry out the exercise as the figure shows, and find that the agreement between a third-order polynomial fit and the VQE result for the calculation of PDM of RaH are 99.96 $\%$ for VQE and 94.57 $\%$ for the CASCI method. An interesting observation is the deviation from a monotonic behaviour of the curves for the PDM in the figure, in the case of MgH.

    
\subsection{Correlation effects}
    \begin{figure}[!t]
        \centering
        \includegraphics[width=0.48\textwidth]{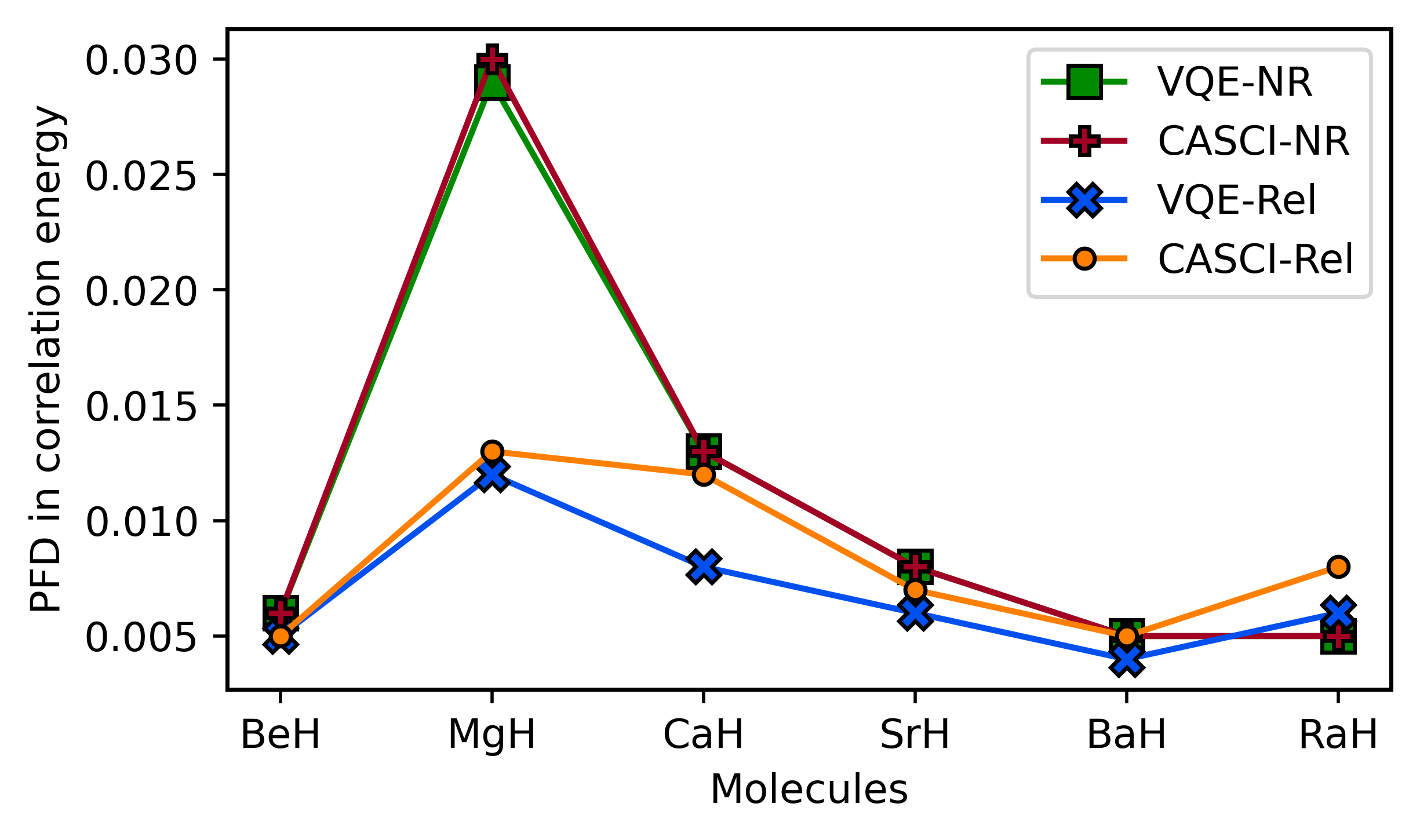}
        \caption{Figure showing the PFD in correlation energy with respect to the total energy at the VQE and CASCI levels.}
        \label{fig:3}
    \end{figure}

We now turn our attention to the role of correlation effects to the ground state energies and PDMs of the considered molecules. For LiH, correlation effects are found to be 0.36 mHa in both the NR and relativistic situations (difference between mean-field and VQE energies as well as the difference between mean-field and CASCI values). For the alkaline earth metal monohydrides, the reported correlation effects at the NR level had a magnitude of 0.97 mHa in 16.732 Ha for BeH to 0.24 mHa in 4.857 Ha (difference between VQE and HF, and the difference between CASCI and HF methods) for RaH. It can be seen from Figure \ref{fig:3} that the percentage fraction difference (PFD) corresponding to these numbers agrees with 96.67 precision for MgH, and with 100 $\%$  precision for all the remaining molecules. For the relativistic case, the correlation effects are 0.87 mHa for BeH (both for VQE and CASCI), whereas it is 0.29 mHa (for VQE) and 0.40 mHa (for CASCI) for RaH. The PFD corresponding to these numbers varies from 0.005 for BeH to 0.008 for RaH. a similar kind of deviation from a monotonic behaviour can also be seen for MgH here. However, the deviation decreases as we calculate the PFD for the relativistic case. Although MgH appears to be anomalous to a large degree for the NR case, we can draw attention to the fact that the difference between the NR and rel cases for MgH is $0.018 \%$.
    
We then conduct an identical set of analyses for the PDMs of the considered molecules. For LiH, correlation reduces the PDM in both NR and rel cases with PFDs of 0.84 for both the VQE and the CASCI methods. In addition, the obtained PDM value from the VQE algorithm is 99.15 $\%$  accurate as compared to the experimental value. For the alkaline earth metal monohydrides, electron correlation reduces the PDM with a PFD of 7.14 $\%$  for BeH (both NR and rel cases; at the VQE and CASCI levels of theory). However, the PFD in correlation energy decreases for the NR case as we move towards heavier molecules for both the VQE and CASCI methods, all the way upto 0.62 $\%$  (for VQE) and 0.31 $\%$  (for CASCI) for RaH. While the behaviour is somewhat different for the rel case, the PFD in correlation energy for both the VQE and CASCI methods decreases till CaH and it increases afterwards.

    \begin{figure}[!h]
        \centering
        \includegraphics[width=0.48\textwidth]{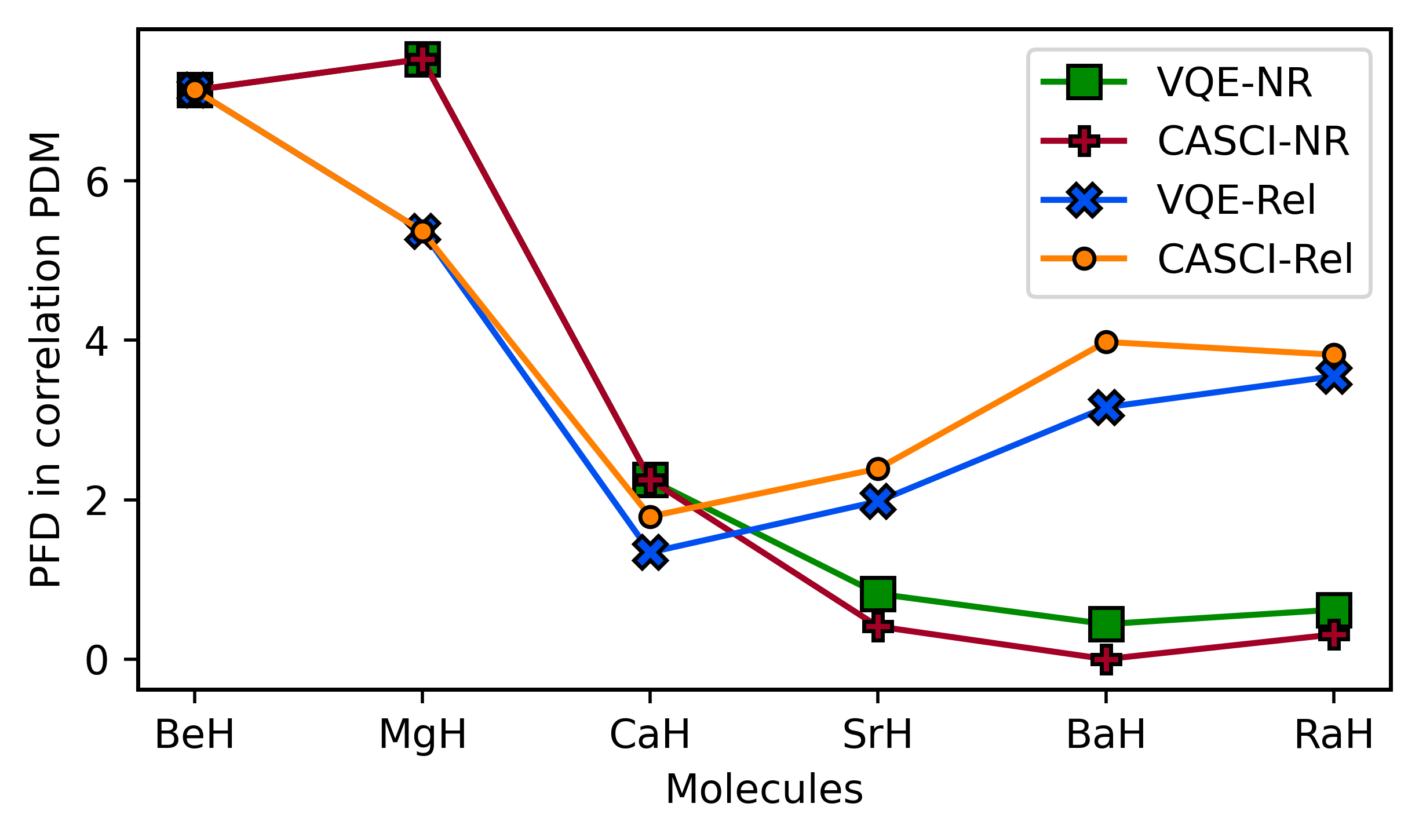}
        \caption{Figure illustrating the PFD in correlation in PDM with respect to the total PDM at the VQE and CASCI levels.}
        \label{fig:4}
    \end{figure}

\subsection{Interplay of relativistic and correlation effects}
Next, we focus on the interplay of both the relativistic and correlation effects to the ground state energies and PDMs. For LiH, the combined relativistic and correlation effects are approximately 1.2 mHa in 8.983 Ha (0.013 $\%$) for energy and 0.05 Debye in 5.93 Debye (0.84 $\%$) for the PDM. For the alkaline earth metal monohydrides, the combined effects of relativity and correlation equal to 3.7 mHa in total of 16.733 Ha for BeH, and 121.415 mHa in 4.736 Ha for RaH, which translates to PFD of 0.022 $\%$ and 2.56 $\%$ respectively. For PDM, the total relativistic and correlation effects add up to 0.02 Debye in 0.28 Debye (7.14 $\%$) for BeH, and 0.74 Debye in 3.94 Debye (18.78 $\%$) for RaH. In the case of the PDM of BaH, the correlation lowers the PDM by $3.16 \%$ after a VQE calculation, while the PFD due to relativistic effects at the mean-field level is approximately $13.41 \%$. As a result, there is a difference of 10.67 percentage overall due to the relativistic and correlation eﬀects. This illustrates the complex nature of the calculations in quantum many-body theory, by virtue of the interplay between relativistic and correlation effects.

\subsection{Sources of error in our calculations}
We now proceed to analyse the factors contributing to errors in our predictions. For this purpose, we have taken SrH as a representative system.

First, we proceed to calculate the errors in our calculations that could stem from the choice of basis. The calculated PFD in energy between the Dyall's v4z and the Dyall's v2z is 0.04 $\%$  and for PDM it is 0.80 $\%$ . Comparatively, the dyall.v4z basis set requires more time to calculate the one- and two-electron integrals, while not altering the results for energies and PDMs noticeably. It is for this reason that we carry out all our calculations in the current work using Dyall's v2z basis. 

Next, we study the impact of choice of active space to our calculations. To that end, we increase the number of virtuals to 14 and 16, thereby increasing the size of the active space of SrH. The obtained PFDs in energy with computations using 14 and 16 qubits are $0.004 \%$ and $0.002 \%$, respectively. On the other hand, the PFD calculated for the PDM is 0.29 $\%$  and 0.23 $\%$ , respectively. It is worth noting at this point that with our choice of active space, the PDMs that we obtain using relativistic VQE agree reasonably well with those from CCSD calculations carried out using a double zeta basis on a traditional computer~\cite{R3,RaH}. The PFD of our relativistic VQE results with respect to the CCSD ones are found to be: BeH: 28 $\%$ , MgH: 21 $\%$ , CaH: 8 $\%$ , SrH: 9 $\%$ , and BaH: 17 $\%$ . Perhaps fortuitously, RaH yields the best accuracy in our results, and its PFD is about 2 $\%$. However, it is also worth noting that while one may expect the core orbitals (for example, $s$ and $p_{1/2}$) to be most affected by relativity, the outer electrons too are influenced considerably for heavy molecules due to the large nuclear charge. Therefore, for a far-nuclear region property such as the PDM, potentially important relativistic effects that contribute to PDM might be effectively captured within our active space for the heavier systems. 

We expect the results to be almost independent of the choice of mapping scheme \cite{m31}, and we hence do not attribute any error due to it. 

We finally comment on our choice of optimizer. We use SLSQP in our work because compared to other state-of-the-art classical optimizers such as COBYLA \cite{6f, 6g} and L-BFGS-B \cite{6h}, it offers the best performance in terms of precision and number of iterations (i.e. time taken). For example, for SrH, SLSQP converges in $\sim550$ seconds, while COBYLA and LBFGSB require $\sim1800$ and $\sim570$ seconds, respectively. The SLSQP also has a lower upper bound for the ground state energy than the other two. \\ 

%% file: sections/section04.tex
\section{Conclusion} \label{sec:conclusions}
We have implemented and employed the relativistic version of the VQE algorithm and applied it to the calculation of ground state energies as well as molecular electric dipole moments of a range of single valence molecular systems, beginning from the lightest BeH to the heaviest RaH of the alkaline earth metal monohydride series. We compare our results with relativistic complete active space configuration interaction (CASCI) approach, as well as with their non-relativistic counterparts in the VQE and CASCI frameworks. We also elucidate the role of correlation as well as relativistic effects, and their interplay in determining the final results. We find from our simulations that the relativistic VQE algorithm is capable of capturing relativistic effects in PDMs with a precision of $ \ge 99.2\%$ for all the considered molecules. We find that our VQE results deviate from the CASCI ones by at most $3.09\times 10 ^{-3} \%$ (for the CaH molecule), while the best agreement can be as low as $4.18\times 10 ^{-5} \%$ percent (in the case of BeH). With our choice of active space, our 12-qubit relativistic VQE computations not only yield very precise results for all the systems considered, but also agrees to within 10 percent for CaH, SrH, and RaH with results obtained with all-electron CCSD calculations on traditional computers. We anticipate that with advancements in quantum hardware, the relativistic VQE algorithm will be able to handle heavier molecular systems and open new avenues for novel applications such as probing new physics beyond the Standard Model of elementary particles. 

%% file: sections/acknowledgements.tex
\section*{Acknowledgements} \label{sec:acknowledgements}
    All our calculations were carried out on the Rudra cluster from SankhyaSutra Labs, Bangalore. K.S. acknowledges support from JST PRESTO "Quantum Software" project (Grant No. JPMJPR1914), Japan and KAKENHI Scientific Research C (21K03407) from JSPS, Japan. BPD, VSP, and KRS acknowledge support from MeitY-AWS Braket QCAL project (N-21/17/2020-NeGD, 2022-24). We would like to acknowldge Dr. V. P. Majety (Indian Institute of Technology Tirupati) for useful discussions with him.

%% file: sections/appendix1.tex
\section{Appendix} \label{sec:appendix}
\setcounter{figure}{0}
\renewcommand{\figurename}{Fig.}
\renewcommand{\thefigure}{S\arabic{figure}}
\begin{figure*}[hbt!]
    \begin{tabular}{c c}
    \includegraphics[scale=0.65]{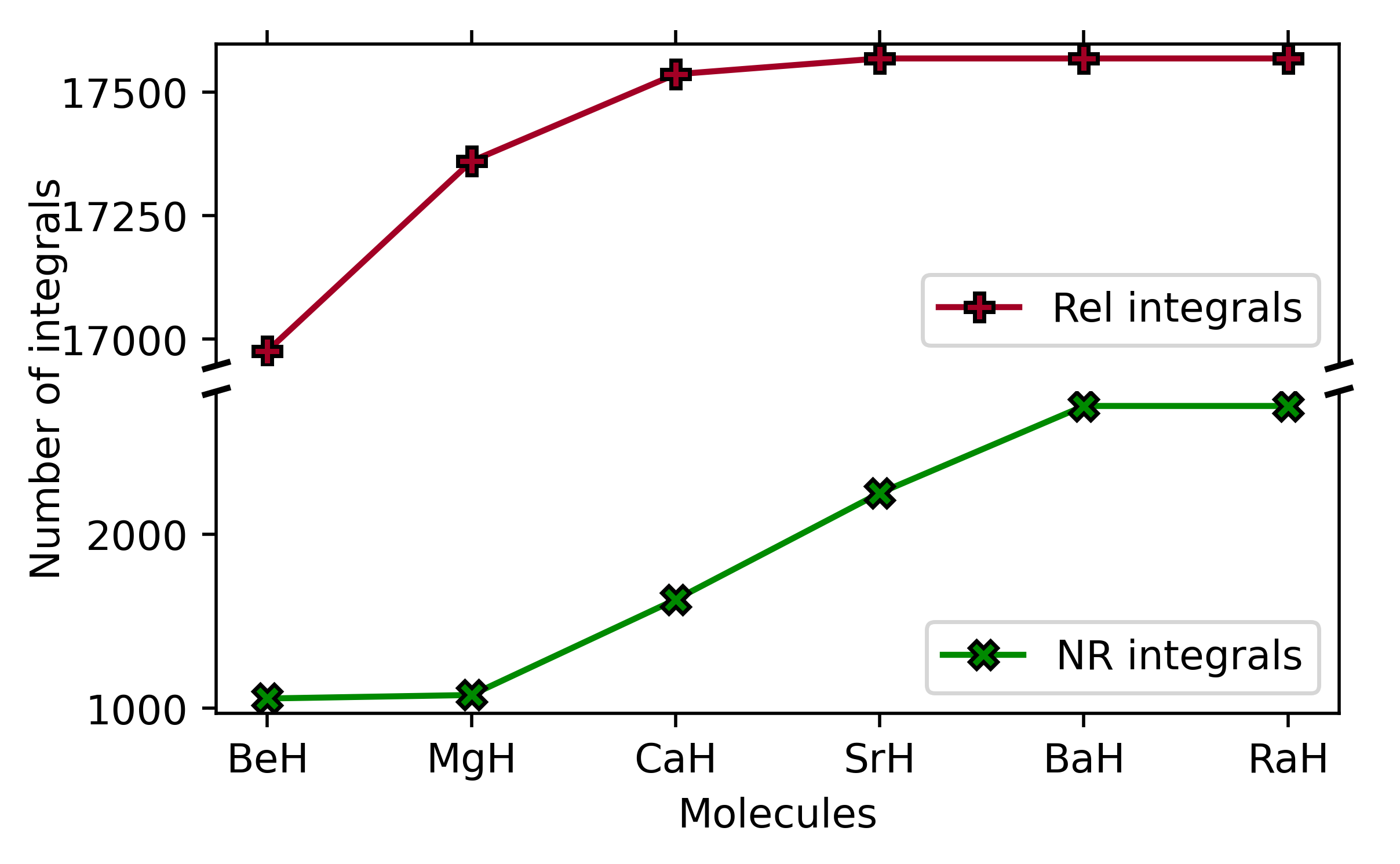} & \includegraphics[scale=0.65]{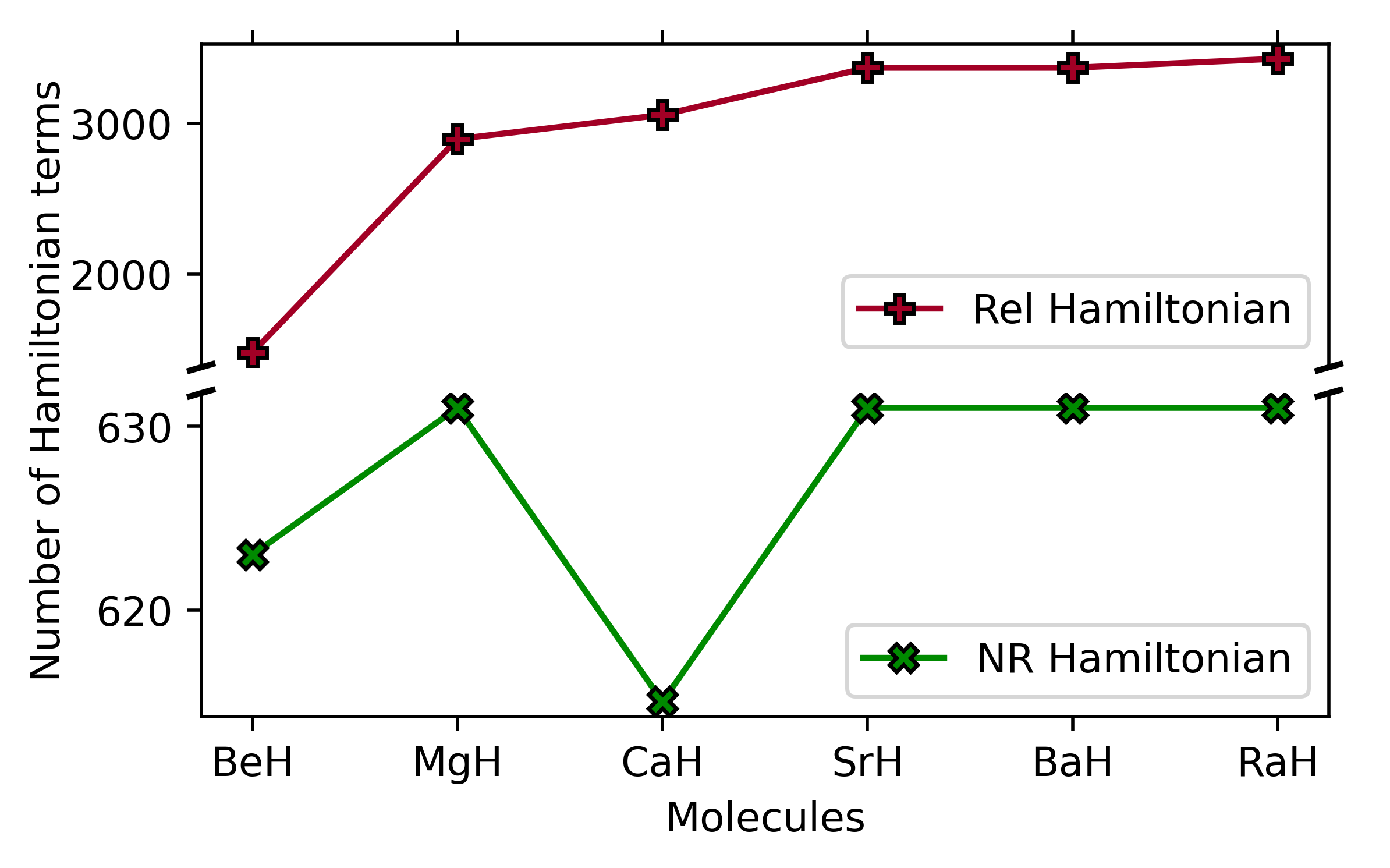} \\
    (a) & (b)\\ \\
    \includegraphics[scale=0.65]{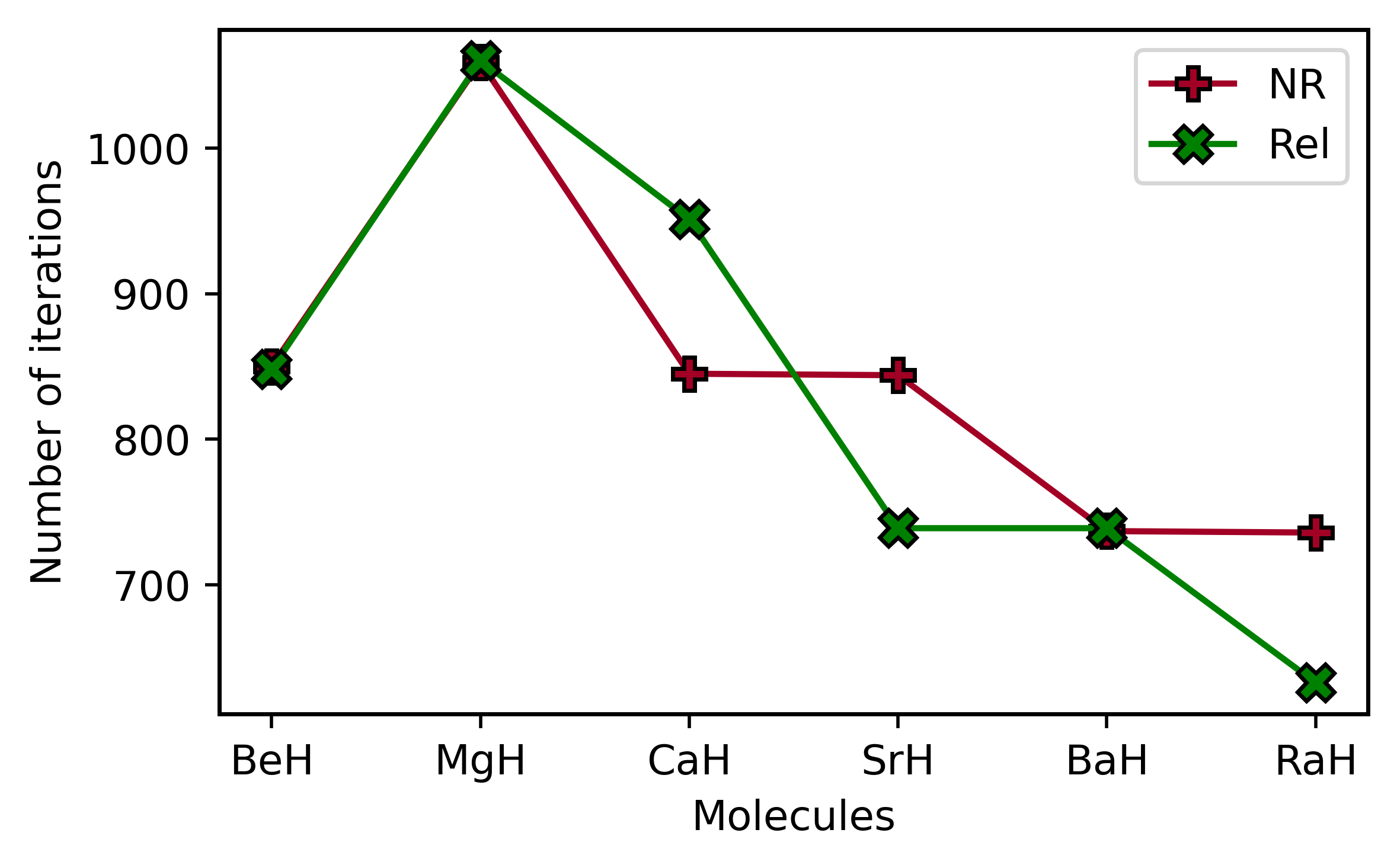} & \includegraphics[scale=0.65]{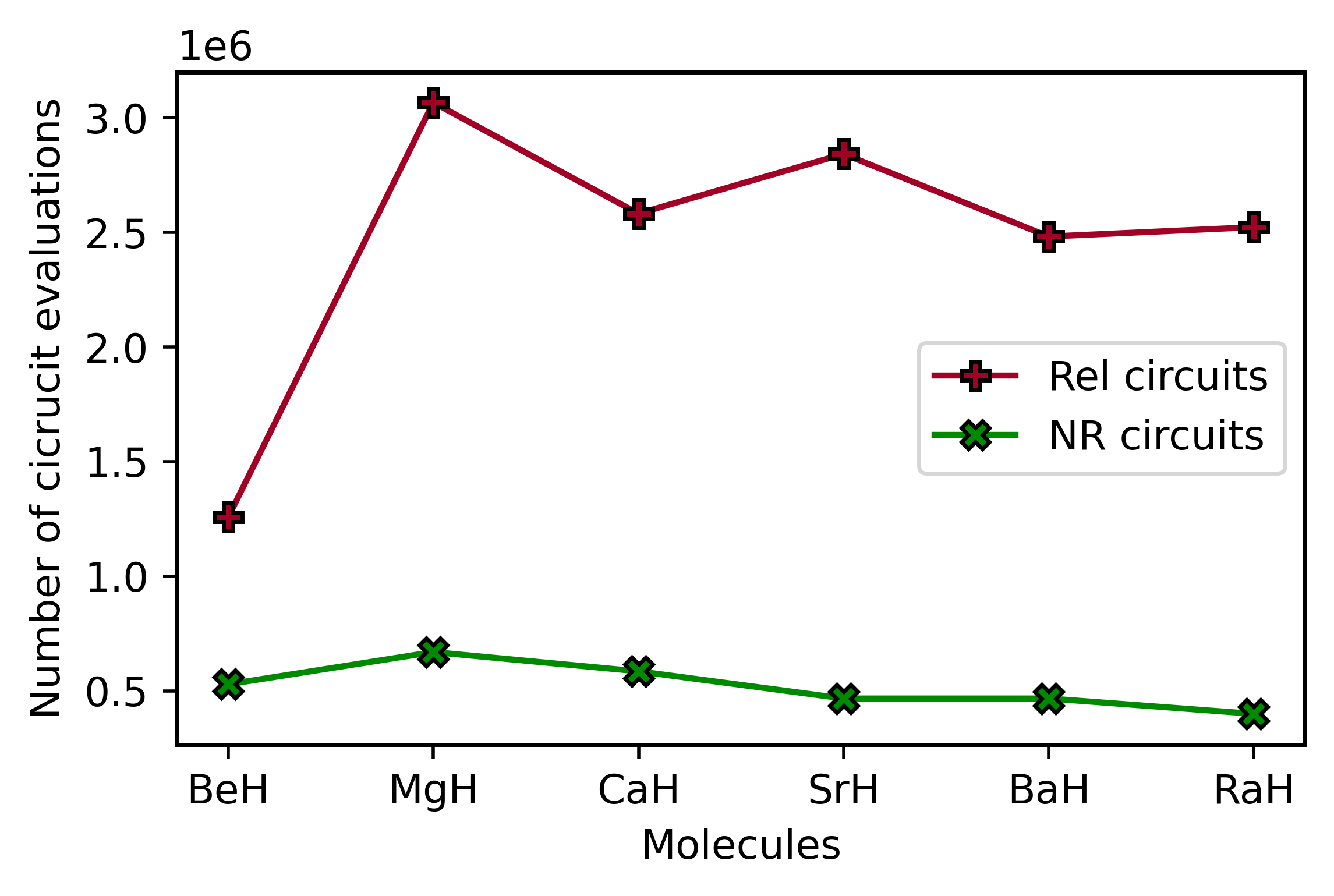} \\
    (c) & (d) \\
    \end{tabular}
    \caption{\label{fig:FIG3} Figure illustrating (a) the total number of one- and two- integrals, (b) the number of terms in the Hamiltonian, (c) the number of iterations needed for convergence, and (d) the number of circuits (in millions) needed to be evaluated, in both the NR and rel cases.} 
\end{figure*}

We have provided plots in the Appendix, as Figure S1 shows, for the total number of integrals (subfigure (a)), the number of terms in the Hamiltonian (subfigure (b)),  the number of iterations required for convergence in energy (subfigure (c)), and finally the number of circuit evaluations (subfigure (d)), for both the non-relativistic (NR) and relativistic (rel) cases, for the considered molecules. The number of integrals is larger for the relativistic case as compared to the NR scenario. This is to be expected because in the non-relativistic case, a large number of matrix elements are zero, such as many of the one-electron integrals of the form, $h_{pq}$, with $p \neq q$. However, in the relativistic case, these matrix elements may not be zero. The Hamiltonian constructed from the NR and rel cases is constrained to be greater than a threshold, which in our case is set to $10^{-5}$. The number of integrals whose values are larger than this threshold are found to increase as we move towards heavier molecules. This is also demonstrated in Fig. S1 (a), where the number of Hamiltonian terms increase as we move towards heavier molecules. This trend is also reflected in Figure S1 (b), where the number of Hamiltonian terms increase as we go to heavier molecules. Fig. S1 (c) shows the number of iterations needed for converge in energy. Finally, we discuss the number of circuits we need to evaluate for each molecule (number of Hamiltonian terms $\times$ number of iterations). For example, the maximum number of circuit evaluations in the rel case is around 3,000,000, whereas in the NR case it is around 650,000 for the MgH molecule. 
    
    \renewcommand{\thetable}{S\arabic{table}}
    \begin{table}[t]
            \begin{ruledtabular}
            \caption{\label{tableS1}
            Table presenting the ground state energies and PDMs of the considered molecules from different methods. The list of abbreviations used are HF: Hartree-Fock, DF: Dirac-Fock, VQE: VQE with UCCSD ansatz,  NR: non-relativistic, Rel: relativistic, CASCI: Full Configuration Interaction. Our main results for this work are marked in bold font. The energy is in units of Hartree, whereas the PDM is given in Debye. 
            }
            \begin{tabular}{cccc}
                \textrm{Molecule}&
                \textrm{Method}&
                \textrm{Energy}&
                \textrm{PDM}\\
                \colrule
                LiH & HF & -8.982092 & 5.98 \\
                 & DF & -8.982902 & 5.98 \\
                 & \textbf{VQE (NR)} & -8.982455 & 5.93 \\
                 & \textbf{VQE (Rel)} & -8.983258 & 5.93 \\
                 & CASCI (NR) & -8.982455 & 5.93 \\
                 & CASCI (R) & -8.983258 & 5.93 \\
                 & Expt. & & 5.88 \cite{LiH_PDM}\\
            \colrule
                BeH & HF & -16.729472 & 0.30 \\
                 & DF & -16.732295 & 0.30 \\
                 & \textbf{VQE (NR)} & -16.730442 & 0.28 \\
                 & \textbf{VQE (Rel)} & -16.733166 &  0.28 \\
                 & CASCI (NR) & -16.730449 & 0.28 \\
                 & CASCI (Rel) & -16.733173 & 0.28 \\
            \colrule
                MgH & HF & -10.059301 & 1.57 \\
                 & DF & -9.977368 & 1.57 \\
                 & \textbf{VQE (NR)} & -10.062254 & 1.46 \\
                 & \textbf{VQE (Rel)} & -9.978534 & 1.49 \\
                 & CASCI (NR) & -10.062282 & 1.46 \\
                 & CASCI (Rel) & -9.978663 & 1.49 \\
            \colrule
                CaH & HF & -6.795599 & 2.27 \\
                 & DF & -6.752359 & 2.27 \\
                 & \textbf{VQE (NR)} & -6.796480 & 2.22 \\
                 & \textbf{VQE (Rel)} & -6.752936 & 2.24 \\
                 & CASCI (NR) & -6.796492 & 2.22 \\
                 & CASCI (Rel) & -6.753145 & 2.23 \\
            \colrule
                SrH & HF & -5.939165 & 2.44 \\
                 & DF & -5.873558 & 2.57 \\
                 & \textbf{VQE (NR)} & -5.939643 & 2.42 \\
                 & \textbf{VQE (Rel)} & -5.873927 & 2.52 \\
                 & CASCI (NR) & -5.939648 & 2.43 \\
                 & CASCI (Rel) & -5.873989 & 2.51 \\
            \colrule
                BaH & HF & -5.234272 & 2.26 \\
                 & DF & -5.149639 & 2.61 \\
                 & \textbf{VQE (NR)} & -5.234535 & 2.25 \\
                 & \textbf{VQE (Rel)} & -5.149859 & 2.53 \\
                 & CASCI (NR) & -5.234536 & 2.26 \\
                 & CASCI (Rel) & -5.149887 & 2.51 \\
            \colrule
                RaH & HF & -4.857260 & 3.20 \\
                 & DF & -4.735551 & 4.08 \\
                 & \textbf{VQE (NR)} & -4.857503 & 3.18 \\
                 & \textbf{VQE (Rel)} & -4.735845 & 3.94 \\
                 & CASCI (NR) & -4.857504 & 3.19 \\
                 & CASCI (Rel) & -4.735954 & 3.93 \\
            \end{tabular}
            \end{ruledtabular}
    \end{table}